# Analysis of WDM-PON for Next-Generation Back- and Fronthaul

Dr. Klaus Grobe and Dr. Jörg-Peter Elbers, ADVA Optical Networking SE, Martinsried, Germany

## Abstract

An analysis of next-generation infrastructure for mobile back- and fronthaul is presented. The same infrastructure can be used for wireline backhaul and dedicated business access. Possible coexistence with FTTH residential access based on NG-PON2 TWDM is also analyzed. Further, different locations for pools of base-band units in fronthaul scenarios are compared with regard to resulting cost. It turns out that fronthaul with highly concentrated base-band unit pools can be cost-efficient even if fronthaul bit rates reach 10 Gb/s.

## Kurzfassung

Eine Analyse verschiedener Infrastruktur-Lösungen für zukünftiges mobiles Fronthaul/Backhaul wird gezeigt. Dieselben Infrastrukturen eignen sich auch für leitungsgebundenes Backhaul und Geschäftskundenanbindungen. Weiterhin wird die Koexistenz mit NG-PON2-basiertem FTTH untersucht. Darüber hinaus werden unterschiedliche Lokationen für die konzentrierten Basisband-Einheiten im Fronthaul hinsichtlich der resultierenden Kosten untersucht. Es ergibt sich, dass Fronthaul mit hochkonzentrierten Basisband-Einheiten selbst bei Bitraten von 10 Gb/s kosteneffizient sein kann.

## 1 Introduction

Beyond 2020, radio-access networks (RAN) will have to support bandwidths that are considerably higher than today. This holds for 4G that already can support advanced radio techniques like Cooperative Multi Point (CoMP), massive Multi-Input Multi-Output (MIMO) antenna configurations and frequency bonding up to 100 MHz radio bandwidth, all of which support higher bandwidths for the mobile User Equipment (UE). It certainly also holds for 5G which intends to clearly exceed 4G in terms of achievable radio bandwidth [1].

Ideally, the RAN should not be a dedicated network but part of a Fixed-Mobile Converged (FMC) network that also supports wireline backhaul (e.g., XDSL backhaul), dedicated broadband business access, and potentially also FTTH or residential optical access. This convergence is required by the pressure toward lower cost. Lower cost is enabled by avoiding any duplicated functionalities, by network function virtualization, and by and better network utilization. Two aspects of convergence can be separated, structural and functional convergence. Structural convergence considers the unified (access transport) network. This is where WDM-PON comes into play, as elaborated hereinafter. Functional convergence considers the unification of functionalities which are required in both, wireline and wireless networks (e.g., authentication).

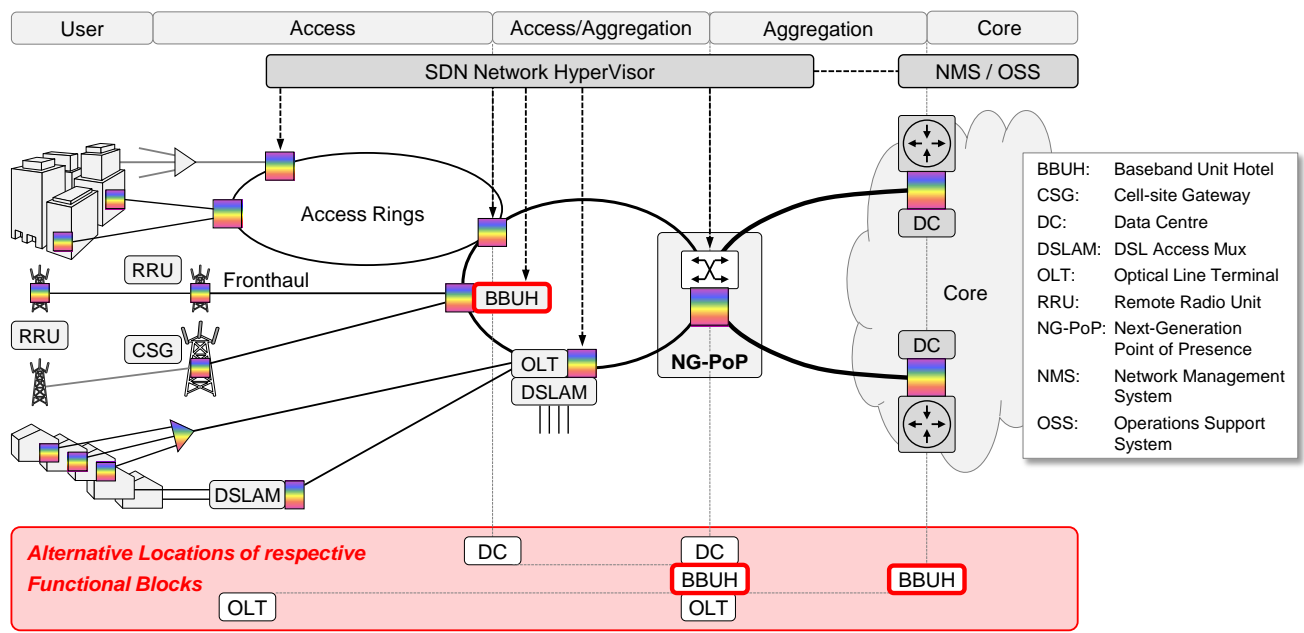

**Figure 1.** Next-Generation Fixed-Mobile Convergence.

Structural convergence must be seen in an accompanying context: active-site consolidation. Here, large network operators (especially ILECs) intend to significantly reduce the number of their active sites (Central Offices, CO) in order to further reduce cost. This leads to two apparent effects. Firstly, mean access distances between the remaining, bigger COs (then called Main CO, MCO) and the clients are increased. These increased distances must be supported by FMC system solutions. Secondly, certain functions in the network, or network elements, can be placed closer toward the clients (e.g., in remaining COs or even in cabinets), or closer toward the core (in core COs, i.e., collocated with core routers etc.). This poses the questions of feasibility and again, cost optimization.

At the time being, it is not clear whether mobile backhaul or fronthaul [2] (or any mix of these or anything in-between that is not yet developed) is most cost-efficient and future-proof. In particular, potentially very high 5G fronthaul bit rates are sometimes mentioned. Therefore, any new converged access infrastructure should have the capability to support such bit rates.

An FMC network example is shown in Figure 1.

## 2   WDM-PON for FMC Networks

The systems solutions for FMC are limited due to the challenging requirements – capacity (including future 5G scaling capability), reach (also considering site consolidation), potential transparency (e.g., for CPRI) [3]. Hence, only fiber-optic solutions which make use of wavelength-division multiplexing and supporting passive infrastructure need to be considered. A similar result was already derived in [4]. By passive infrastructure, an infrastructure without active fan-out elements like switches or routers is meant. This helps consolidating the active aggregation toward fewer levels and less sites. Passiveness also supports minimum energy consumption, which has been shown, e.g., in [5], [6].

Today, passive WDM, in the form of CWDM [7], is used for many backhaul applications. CWDM lacks some scaling capability, and no developments to bit rates exceeding 10 Gb/s are known. However, it can be regarded the reference system solution for any newer solutions. These solutions are NG-PON2 [8], [9] and more general variants of passive DWDM or DWDM-PON (i.e., wavelength-multiplexed PON which are not fully compliant with the NG-PON2 recommendations). Overviews on WDM-PON can be found in [10], [11]. These systems can, e.g., comply with the ITU-T Recommendations G.698.1 [12], G.698.2 [13], G.9802 (former G.multi) [14], or the upcoming Recommendation G.metro [15]. These systems are analyzed hereinafter.

A relevant question for all WDM-PON (which does include NG-PON2) relates to the Optical Distribution Network (ODN, i.e., the passive fiber plant). The ODN for multi-wavelength PON can be based on power splitters or wavelength filters. Systems that support power-split ODN are referred to as Wavelength-Selective (WS-) WDM-PON, those that require filters are called Wavelength-Routed (WR-) WDM-PON. WS-WDM-PON can support legacy ODN and consequently FTTH. WR-WDM-PON does not suffer from the high power-splitter insertion loss and certain crosstalk effects [16], [17].

WR- and WS-WDM-PON in front- and backhaul applications are contrasted in Figure 2.

An extensive analysis regarding resulting total capital expenditures (CapEx) for the different FMC cases (wireline plus mobile backhaul or fronthaul) and system solutions has been conducted in the EU FP7 project COMBO [18]. Here, different traffic-growth scenarios were considered.

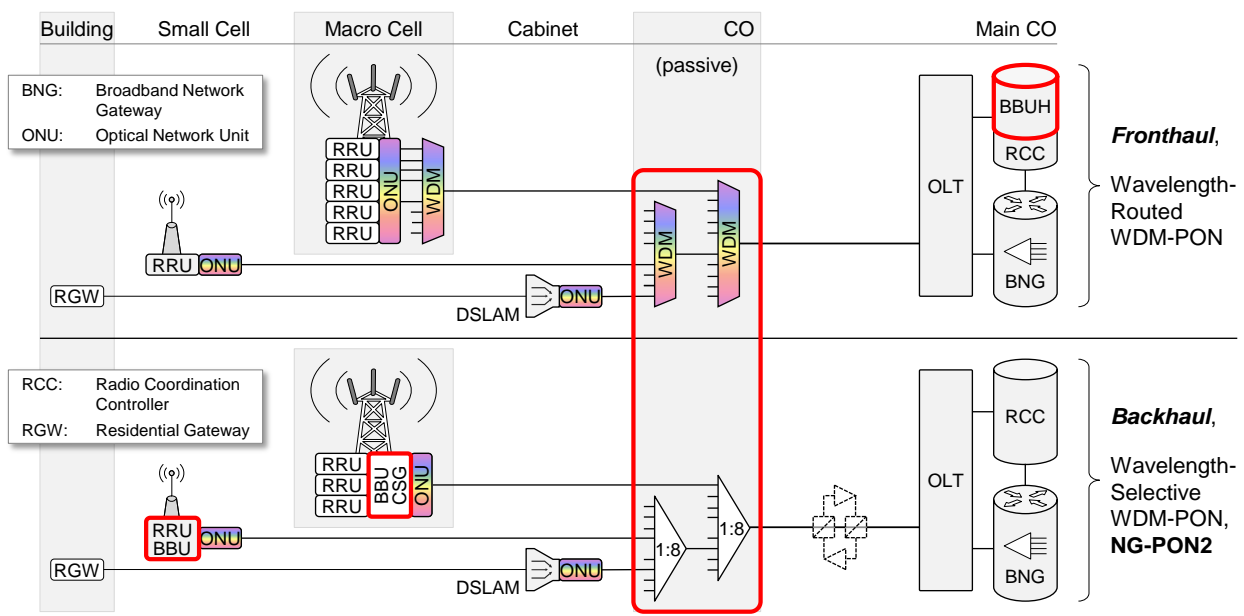

**Figure 2.** WR-WDM-PON for fronthaul (top) vs. WS-WDM-PON / NG-PON2 for backhaul (bottom).

For mobile traffic growth, this translates to varying numbers of newly to be installed small cells. The respective analysis was done for different fiber roll-out assumptions (i.e., assuming FTTC or FTTH areas) and for the different network geo types dense-urban, urban, suburban, and rural. Key differences can be derived between the systems solutions, and for backhaul vs. fronthaul, respectively. Most notably, the number of (optical) interfaces is significantly higher in fronthaul, compared to backhaul. The higher interface number results from the need to connect individual Remote Radio Units (RRU), instead of Base-Band Units (BBUs, which can in turn support multiple RRUs). It also leads to somewhat higher numbers of passive components (filters, power splitters where applicable) and shelves, respectively. In addition, per-channel bit rates are also higher in fronthaul.

There are also differences regarding total fiber length. Fronthaul requires slightly more fiber compared to backhaul. This effect is weak because the same sites have to be connected. The small remaining difference is an effect of the higher fronthaul channel number.

The other relevant difference between backhaul and fronthaul refers to total system latency. Under the assumption of standard Layer-2 aggregation in the backhaul Optical Line Terminations (OLTs, i.e., no special low-latency switching), this difference is in the range of 10 µs and must be considered in the context of advanced radio techniques like Cooperative Multi Point. However, this difference must also be put into relation with absolute fiber delay (run time), which can be as high as up to 400 µs.

On the system level, CWDM and WR-WDM-PON do not require reach extenders (amplifiers) in the field, which helps lowering system cost. This is due to the fact that they make use of wavelength filtering with low insertion loss.

Additional fiber-length differences result from the systems solutions. Under similar assumptions regarding fiber roll-out (e.g., in FTTC or FTTH areas), there are advantages for NG-PON2 because of its fiber reuse capability. Slight advantages also result for WDM-PON with high channel count (i.e., fiber capacity).

A fully converged solution, on fiber and system level and including FTTH, can only be achieved with NG-PON2 on power-split ODN with additional PtP WDM PON overlay. For cases where (re-) usage of power-split ODN is no strict requirement, WR-WDM-PON shows certain advantages in that it avoids any amplifiers in urban areas, and field-deployed amplifiers in rural areas. In addition, there are minor advantages with regard to total fiber length. The latter is driven by the maximum possible channel number. From that, one can conclude that for infrastructure deployments, the number of wavelength channels per system and passive system reach should be as high as possible.

Table 1 gives a summary of the system comparison.

**Table 1.** Comparison of infrastructure options.

|  | CWDM | NG-PON2 | WR-WDM PON | WS-WDM PON |
|---|---|---|---|---|
| Fiber Demand | ● | ●●● | ●● | ●● |
| Passive Reach | ●●● | ● | ●● | ● |
| Capacity / Fiber | ● | ● | ●●● | ●● |
| Infrastructure Reuse | ●● | ●●● | ● | ●● |

The resulting cost comparison between the most relevant system solutions, NG-PON2 and WR-WDM-PON, is shown in Figure 3 [18]. This comparison holds for urban areas, but similar results are obtained for other geo types as well. Two relevant parameters are the fiber availability (fiber-poor vs. fiber-rich areas) and the related FTTH coverage (take-rate, i.e., clients connected), and the number of small cells. The latter indicates the intended mobile-bandwidth growth, it is given as the number of small cells per Macro Base-Station (SC per MBS). The figure shows the resulting trajectories where cost of NG-PON2 and WR-WDM-PON is identical. These curves separate the areas (in dependence on fiber availability/usage and small-cell number) where the one or the other solution leads to better cost. In general, WR-WDM-PON becomes more competitive for lower small-cell numbers and in particular for areas with high fiber availability.

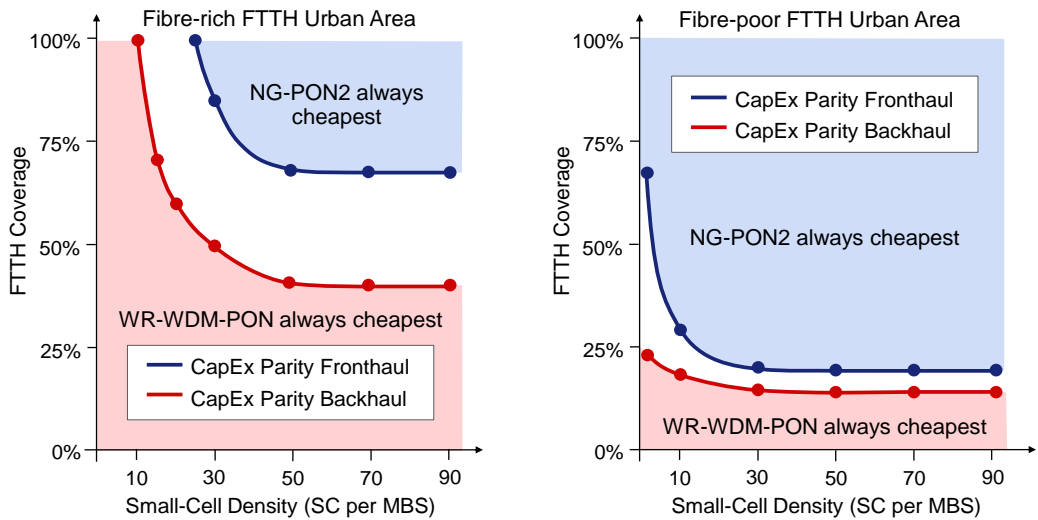

**Figure 3.** Comparison between NG-PON2 and WR-WDM-PON in urban areas for different fiber coverage, fiber availability and small-cell numbers.

## 3  BBUH Placement

So far, the analysis revealed that fronthaul is somewhat more costly than backhaul, due to the requirements for more transport channels and higher bit rates. However, fronthaul also enables increased radio bandwidth (through tighter control of latencies and better support of radio-coordination techniques). It further allows the separation of the BBUs from the antennas (or RRUs). This in turn allows concentration of several BBUs in common sites and consequently, different aspects of pooling gain. One gain aspect refers to CapEx. The number of BBUs in a pool (which is mostly referred to as BBU Hotel, BBUH) is smaller than the number of dedicated, distributed BBUs. This is enabled by pool oversubscription and by the so-called tidal effect [19]. Less BBUs also consume significantly less energy, thus saving operational expenditures (OpEx). In addition to this, common equipment in the pools or BBUHs – power supply, controllers – can be made significantly more efficient. This again saves CapEx and OpEx as has been shown in [20]. Total CapEx saving of BBU pooling can exceed 20%, whereas total energy-cost saving in suitably large BBUHs, including energy saving modes can be in the range of 50%. The resulting total-cost reduction can slightly overcompensate the higher CapEx of fronthaul (with significant pooling gain), compared to backhaul where pooling gain is not possible. This holds for fronthaul bit rates up to the range of 10 Gb/s.

The pooling gain depends on the pool or BBUH size, that is, on the possibility for oversubscription and statistical multiplexing. These effects get better the bigger the pools are, i.e., the more they are located in very few sites, toward or in the core of the network. However, analysis in [18] also revealed that within the maximum fronthaul latency allowed, only 35…55% of the antennas could be served from the Core COs. From the Main COs, this increases to >95% which, in practice, allows a unified solution. (Rare cases of higher rural distances can be served with dedicated fiber solutions.) Below the MCOs, in remaining local COs or cabinets, pooling gain is smaller due to insufficient statistical multiplexing. Therefore, BBUH accommodation in the MCO is recommendable, and the pooling-gain numbers stated earlier hold for this placement.

## 4  Bit Rates Higher than 10 Gb/s

Fronthaul in 4G can reach bit rates of 10 Gb/s per sector for 20 MHz radio bandwidth and MIMO degree of 8. Theoretically, up to 100 MHz radio bandwidth can be used, leading to 50 Gb/s fronthaul bit rate per sector. Today, this is regarded an unlikely scenario due to bandwidth availability. These bit rates assume 8B/10B line coding. With 64B/66B coding, the higher bit rate can be reduced into the range of 41 Gb/s.

5G intends to use even higher bandwidth, and possibly higher MIMO degrees. However, the question remains if in particular the higher bandwidth will be available on large scale. Therefore, it is questionable if fronthaul bit rates clearly exceeding 40 Gb/s will become necessary in 5G. In addition, next-generation fronthaul may use techniques for bit-rate reduction. Then it should be shown for guaranteed future that bit rates in the range of 28…43 Gb/s are feasible at suitably low cost.

This has been analyzed in [21]-[24]. Both, low-cost potential and reach capability suitable for the site-consolidated access scenario described before have been investigated. Contending modulation schemes for these bit rates with low-cost potential included direct-detection PAM4 and optical and electrical Duobinary. Cost mark-up of the resulting transceivers, measured against 10-Gb/s WDM pluggable transceivers, was estimated under the assumption of fully ramped-up mass production. In this analysis, mark-up factors of ~2 for 28 Gb/s and ~3 for 40 Gb/s were found. This indicates that bit rates of 28…40 Gb/s can have reasonably low cost, since the transceivers are only part of the total cost. Other components (e.g., housing, controllers, filters, connectors) are not affected by bit-rate increase. Therefore, the mark-up for complete links is lower than the factors mentioned before.

Achievable reach for direct-detection schemes with 28 Gb/s and 40 Gb/s was also investigated, including first experimental validation [23], [24]. For 28 Gb/s, reach of up to 60 km for PAM4 and 40 km for electrical Duobinary is achievable, respectively. For 40 Gb/s, reach in the range of 30 km is achievable for PAM4, and 20 km for electrical Duobinary. These values include the use of optical amplifiers in the respective PON OLTs. Optical Duobinary has slightly shorter reach, compared to electrical Duobinary. However, for that reach, lower receive power is required. For the respective reach domain (<15 km), this allows somewhat simpler receivers. On the other hand, Optical Duobinary requires a somewhat more complicated modulator. Hence its use in high-speed access is questionable. In general, with increasing bit rates, different modulation schemes achieve lower cost or better sensitivity in different reach domains. Therefore, for high bit rates, different solutions may have to be offered for different reach requirements for cost optimization.

## Conclusion

Next-generation access networks require a converged infrastructure for fixed and mobile access. At the same time, mean access distances are increasing due to ongoing attempts of active-site consolidation. The access infrastructure must therefore support long reach up to the range of 50 km and high accumulated and per-channel capacities. This can be achieved cost-efficiently with WDM-PON, including NG-PON2. A more detailed analysis reveals that Wavelength-Routed WDM-PON and shared-spectrum NG-PON2 (i.e., the combination of TWDM and PtP WDM PON overlay) are the most effective solutions for converged and site-consolidated access. The choice between them depends on fiber availability and the number of connections (to small cells) to be provided. For high ratio of fibers/connections, WR-WDM-PON is more advantageous, for smaller ratios, NG-PON2 is the best choice.

When fronthaul and backhaul are compared, slightly higher cost for the fronthaul infrastructure results. However, this can be compensated by pooling gain in BBUHs, leading to even slightly lower total cost for fronthaul. This holds for fronthaul bit rates up to ~10 Gb/s. For even higher bit rates (which are possible in 5G next-generation fronthaul), feasibility for low-cost implementation of bit rates at least in the range of 28…40 Gb/s is relevant. Here, first analysis showed that transceivers for this bit-rate range can have cost that is higher by factors 2…3 compared to 10-Gb/s transceivers. The respective implementations also have reach capabilities sufficient for converged access infrastructure.

Therefore, it can be concluded that next-generation converged access is feasible even for very high future bit rates at reasonably low cost. The underlying infrastructure will be based on WDM-PON.

## Acknowledgement

The work leading to these results has received funding from the European Union's seventh Framework Program (FP7 2007/2013) under Grant Agreement n° 317762 (ICT-COMBO), and from the European Union's Horizon 2020 Research and Innovation Program under Grant Agreement n° 644526 (iCIRRUS).

## Biographies

**Klaus Grobe** is Director Global Sustainability at ADVA Optical Networking. He received the Dipl.-Ing. and Dr.-Ing. degrees in electrical engineering from Leibniz University, Hannover, Germany, in 1990 and 1998, respectively.  Klaus is one of the authors of Wavelength Division Multiplexing – A Practical Engineering Guide (Hoboken, NJ, Wiley, 2014) and authored and co-authored more than 100 technical publications, three further book chapters and 30 (pending) patents. He is senior member of the IEEE Photonics Society and member of VDE/ITG and ITG Study Group 5.3.3 on Photonic Networks. He represents ADVA in FSAN and QuEST Forum.

**Jörg-Peter Elbers** is Vice President Advanced Technology at ADVA Optical Networking. He received his diploma and Dr.-Ing. degrees in electrical engineering from Dortmund University, Germany, in 1996 and 2000, respectively. Jörg authored and co-authored more than 100 technical presentations, 3 book chapters and 15 patents. He is member of the IEEE and VDE/ITG as well as OFC program committee member and associate editor of JOCN. Jörg also serves on the board the European Technology Platform on Photonics and the European 5G Infrastructure Association.